\documentclass[11pt]{article}
\usepackage{authblk}
\usepackage{graphicx}
\usepackage{graphicx}
\usepackage{epstopdf}
\newcommand{\celsius}{\,^{\circ}{\rm C}}
\usepackage[margin=1in]{geometry}
\title{Strand displacement: a fundamental mechanism in RNA biology?}

\author[1]{Fan Hong}
\author[1,2,*]{Petr \v{S}ulc}
\affil[1]{\footnotesize{Center for Molecular Design and Biomimetics at the Biodesign Institute and School of Molecular Sciences, Arizona State University, Tempe, Arizona 85287, United States}}
\affil[2]{Center for Biological Physics, Arizona State University, Tempe, AZ, 85287-1504, USA}
\affil[*]{E-mail: psulc@asu.edu}
\date{}

\begin{document}
\maketitle

\begin{abstract}

DNA and RNA are generally regarded as central molecules in molecular biology. 
Recent advancements in the field of DNA/RNA nanotechnology successfully used DNA/RNA as programmable molecules to construct molecular machines and nanostructures with predefined shapes and functions. The key mechanism for dynamic control of the conformations of these DNA/RNA nanodevices is a reaction called strand displacement, in which one strand in a formed duplex is replaced by a third invading strand. While DNA/RNA strand displacement has mainly been used to \textit{de-novo} design molecular devices, we argue in this review that this reaction is also likely to play a key role in multiple cellular events such as gene recombination, CRISPR-based genome editing, and RNA cotranscriptional folding. We introduce the general mechanism of strand displacement reaction, give examples of its use in the construction of molecular machines, and finally review natural processes having characteristic which suggest that strand displacement is occurring.  

\end{abstract}
\section{Introduction}


DNA and RNA molecules are key molecules in all living systems. While DNA is used exclusively for information storage, RNA is a versatile molecule involved in multiple cellular processes that include information storage and transfer, catalysis, and genetic regulation. 


From an engineering perspective, nucleic acids are among the most programmable molecules because of their relatively predictable interactions and affordable synthesis process. In the past two decades, researchers started to exploit these advantages to dynamically control interactions between nucleic acid strands to build artificial molecular machines with designed kinetic and thermodynamic behavior. Strand displacement (sometimes also referred to as strand exchange or branch migration in these cases) is the key process of such dynamic control. 
Strand displacement occurs when two nucleic acid strands hybridize,  while displacing a strand that was previously bound to one of them (illustrated in Fig.~\ref{fig:displacement} and described in detail in Section \ref{sec:dispmechanism}). The reaction does not require any external driver, such as helicases to unwind the initial duplex. 
 The mechanism works extremely well in both \textit{in vitro} and \textit{in vivo} systems. For example, researchers are able use DNA or RNA to build complex biochemical circuits \cite{srinivas2017enzyme,yin2008programming}, molecular computers \cite{qian2011neural,cherry2018scaling,qian2011scaling}, and cellular machines to control gene expression \cite{chappell2015creating,isaacs2004engineered,green2014toehold,mutalik2012rationally,green2017complex}. 
%
Given the ability of RNA and DNA molecules to perform the strand displacement reaction, and the success in numerous engineering applications of this cascade, it is of interest to ask if the same mechanism is also exploited by natural systems. 
In fact, in several studies, authors have explained plausible rearrangement of interacting nucleic acids that is highly reminiscent of strand displacement.

This review is organized as follows. We first introduce the mechanism of strand displacement and review some engineered molecular systems and cellular machines utilizing RNA or DNA strand displacement in nanotechnology and synthetic biology. We then review several nucleic acid interactions in nature that likely involve strand displacement reactions. We discuss the role of strand displacement reaction in DNA replication fork, CRISPR-Cas systems, rearrangement during cotranscriptional RNA folding, spliceosome assembly, and other systems that undergo rearrangement of competing domains such as riboswitches. 




 
\section{The mechanism of strand displacement}
\label{sec:dispmechanism}

\subsection{Features of strand displacement kinetics}
\begin{figure}
  \includegraphics[width=1.0\textwidth]{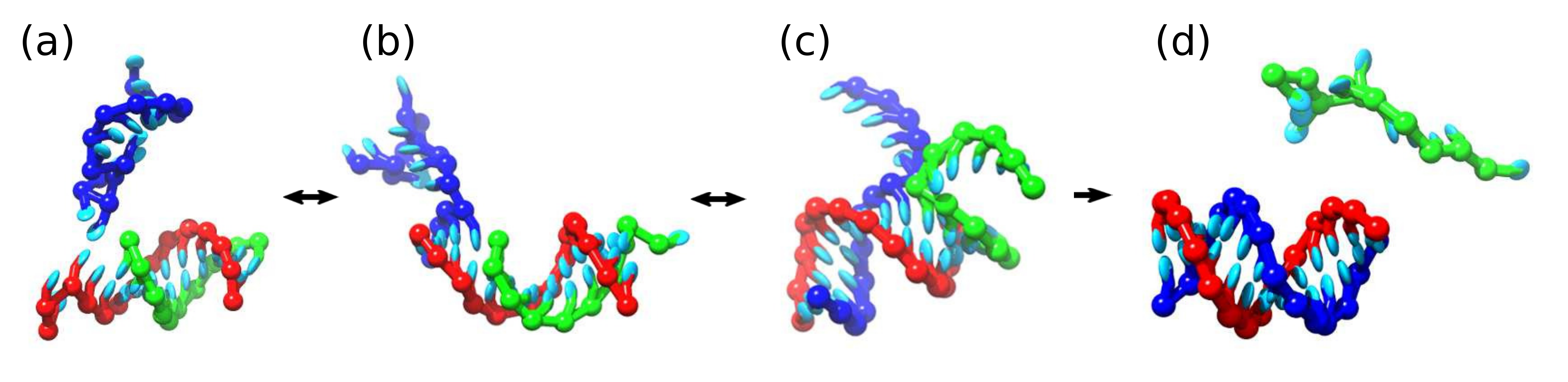}
  \caption{A schematic representation of toehold-mediated strand displacement in a three-strand RNA system, as represented in a computer simulation with a coarse-grained model of RNA. 
   (a) The invading (blue) strand attaches to the single-stranded region (called the toehold) on the substrate (red) strand, which was previously paired with the incumbent (green) strand. (b) The invading strand is fully bound to the toehold. (c) The invading strand exchanges bases with the incumbent strand. (d) The invading strand eventually displaces the incumbent strand and fully binds to the substrate strand.}
  \label{fig:displacement}
\end{figure}

\begin{figure}
\centering
  \includegraphics[width=1\textwidth]{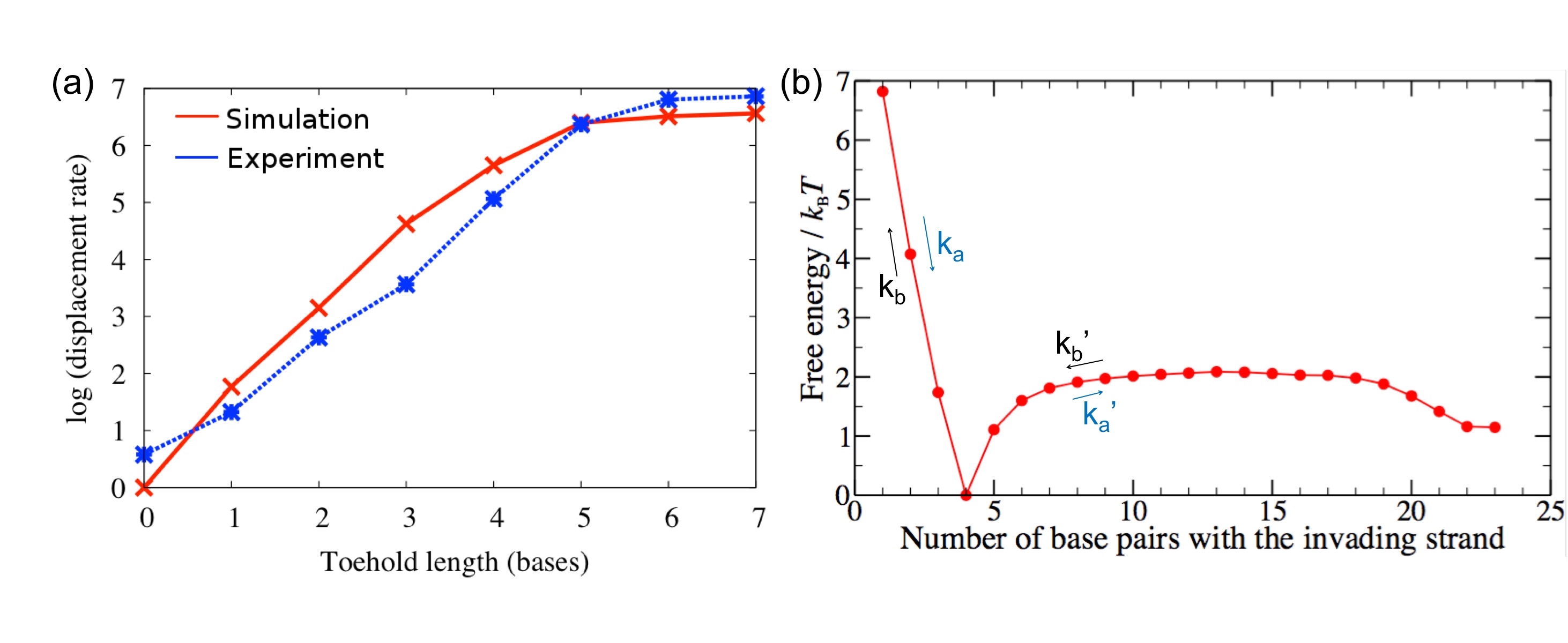}
  \caption{(a) Measurement of DNA strand displacement rate (relative to the rate measured for toehold length 4) 
   as a function of toehold length in experiment from Ref.~\cite{zhang2009control} (red line) and in simulation from Ref.~\cite{Displacement} (blue dashed line) at $25\celsius$.  (b) Free-energy profile of the displacement reaction with toehold of length 4.}
  \label{fig:dispkinetics}
  
\end{figure}

%
%

The strand displacement reaction is often used in engineered nanotechnological systems, which we will discuss in Section \ref{sec:nanotech}. It has been experimentally demonstrated for both DNA and RNA, 
and has been also been subject to computational studies 
\cite{Displacement,vsulc2015modelling}. The strand displacement (illustrated in Fig.~\ref{fig:displacement}) refers to a process where an ``invading'' single strand competes for binding to a complementary strand (``substrate'') with another strand (``incumbent'') that has been previously bound to the same ``substrate'' strand. 
  Strand displacement provides a mechanism for exchanging large duplexes. The free-energy barrier to the spontaneous dissociation of longer incumbent strands can be too prohibitively large to happen at experimental timescales; however, spontaneous fraying of single 
base pairs is common, and allows the invading strand gain a base pair with the substrate that was previously bound to he incumbent 
 strand. While we illustrate strand displacement on a three-strand 
system in Fig.~\ref{fig:displacement}, it is also possible to have two of the domains on the same strand (for instance when an invading strand opens a hairpin stem) or all three domains on the same strand. 

During the strand displacement, the invader and incumbent exchange base pair with the substrate. The process can, of course, proceed in both ways when the incumbent strand 
regains base pairs it previously lost to the invader, representing the flat region in Fig.~\ref{fig:displacement}b. In most engineering applications, the invader strand is also complementary to a single-stranded region (called the 
``toehold'') immediately next to the displaced domain. Therefore, once the incumbent strand is displaced, the invader-substrate duplex will have more base-pairs than 
the original incumbent-substrate duplex, making the backward process less favorable. Such a setup is usually referred to as ``toehold-mediated'' strand displacement. When no toehold is present, strand displacement is still possible, as the end base-pairs of a duplex fray, presenting a unpaired base on the substrate strand that the invader can bind to, initiating the displacement reaction.

Due to the prevalent use of DNA in nucleic acid nanotechnology, the experimental efforts on quantifying the kinetics of strand displacement was focused on DNA strand systems.  
The rate of the strand displacement reaction as a function of the number of nucleotides was investigated both experimentally \cite{zhang2009control} and in simulation 
\cite{Displacement}, and is shown in Fig.~\ref{fig:dispkinetics}. In the experiment and in the simulation, an average-strength (in terms of having approximately the same number of CG and AT base pairs) 
was used in order not to distort the effect by varying sequence stability. 

As seen in Fig.~\ref{fig:dispkinetics}, the rate of displacement increases by about an order of magnitude with each extra base-pair in the toehold region, up to a saturation point of 6-7 toehold bases. This sensitivity to the toehold length can be rationalized through the free-energy profile (Fig.~\ref{fig:dispkinetics}b). Once the invader is fully attached to the toehold, it can either dissociate or proceed to fully displace the incumbent strand. When the toehold becomes long enough (about 7 bases for the average strength toehold studied in Fig.~\ref{fig:dispkinetics}), the barrier to spontaneous detachment becomes too large and the detachment of the invader 
is extremely unlikely. Therefore, after the full attachment of the invader to the toehold, the system eventually proceeds to full displacement of the incumbent, and the rate of the reaction is only limited by the time it takes to diffuse and fully bind to the toehold of the substrate. 
 The experiments shown in Fig.~\ref{fig:dispkinetics} were performed in very diluted system ($\approx$ nM concentrations of the invader and the substrate), so the time to displace 20 bases long substrate is shorter compared to the diffusion time of the invader. 

Another important feature of the displacement is that the branch-migration process is not isoenergetic, and contains an approximately $2\,k_{\rm B} T$ barrier in the free-energy profile for the onset of branch migration (Fig.~\ref{fig:dispkinetics}), where the incumbent strand loses base-pairs to the invading strand. The barrier can be rationalized through steric interference of the single-stranded overhangs at the junction, which lead to reduced conformational freedom as the overhangs 
of the invading and incumbent strands clash in the space constrained by the already-paired regions. This free-energy barrier was confirmed in experiment with DNA strands \cite{Displacement}. 
Furthermore, the transitions in the strand exchange between the invader and the incumbent are slower than the rate of fraying/zippering of the base-pair, due to longer-lived metastable states. In order to model the kinetics of this process accurately, different rates need to be used for the branch migration process ($k^{\prime}$ in Fig.~\ref{fig:dispkinetics}) than for 
fraying. The duration of a single branch migration step is estimated to be approximately $10$-$100 \, \mu {\rm s}$ based on experimental measurements 
\cite{green1981reassociation,radding1977uptake} and stochastic kinetics modeling \cite{Displacement}.

An approximate empirical model of strand displacement has been derived in Ref.~\cite{Displacement}. For the case of long toehold (saturated regime), the strand displacement rate is given by the rate of the hybridization of the toehold region. This rate is weakly sequence-dependent, as the toehold with higher stability has higher probability to fully form after the initial binding \cite{ouldridge2013dna}. On the  other hand, for short toeholds, the rate of strand displacement is proportional to $\exp \left(-\Delta G_t /k_{\rm B} T \right) / b$, where $b$ is the length of the incumbent strand bound to the substrate, and $\Delta G_t$ is the free energy of the fully formed toehold.

\begin{figure}
\centering
  \includegraphics[width=1.0\textwidth]{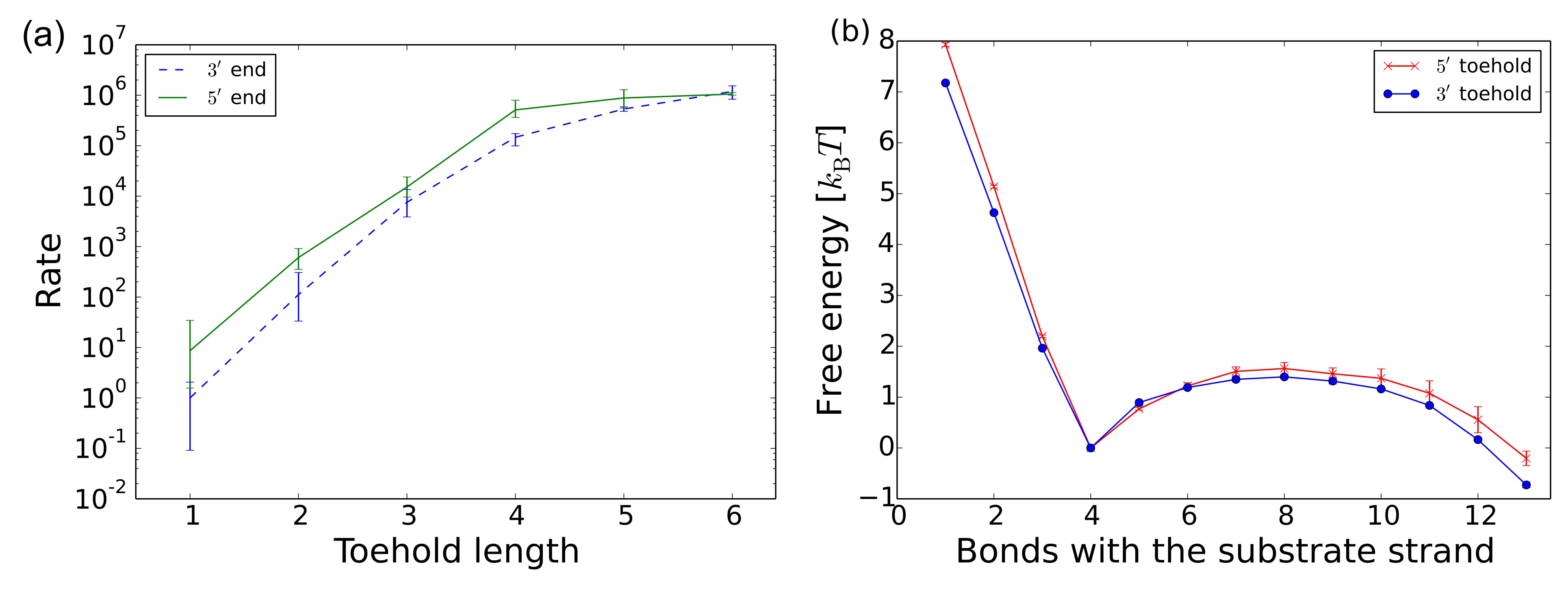}
  \caption{(a) Measurement of RNA strand displacement relative rate at $37 \celsius$ as a function of toehold length with oxRNA model simulation (b) Free-energy profile of the displacement reaction with toehold of length 4 for an invading RNA $5^{\prime}$ and $3^{\prime}$ end.}
  \label{fig:rnadisplacement}  
\end{figure}

The available experimental studies of the branch migration have focused mostly on the DNA strand displacement. The process of RNA strand displacement has been addressed through computational modeling with a coarse-grained model of RNA, oxRNA, in Ref.~\cite{vsulc2015modelling}, which has shown qualitatively the same behavior as the strand displacement kinetics for DNA (illustrated in Fig.~\ref{fig:rnadisplacement} for the averaged-strength oxRNA model, where the interactions between complementary Watson-Crick base-pairs are set to the mean of the interaction strengths). 
 In contrast with the DNA strand displacement system, 
  the simulations of RNA showed higher rates of strand displacement when toehold was placed on the $5^{\prime}$ end, originating from the extra stabilization of the toehold-bound state due to cross-stacking 
interaction between the invader and the substrate, which is only available if the the invader binds to the toehold $5^{\prime}$ end. Simulations have showed that
this effect can lead up to a 10 fold speed-up. However, this effect is only present for toeholds shorter than the saturation length.

Currently, systematic experimental studies of RNA strand displacement kinetics are not available. Based on the simulation results, it is expected that similar timescales and behaviors observed for DNA-strand displacement cascades hold for RNA as well. Synthetic biology/RNA nanotechnology experiments that employ RNA strand displacement (reviewed in Section \ref{sec:nanotech}) designed based on knowledge of similar DNA cascades have so far proven to be successful and functional.

\begin{figure}
\centering
  \includegraphics[width=0.8\textwidth]{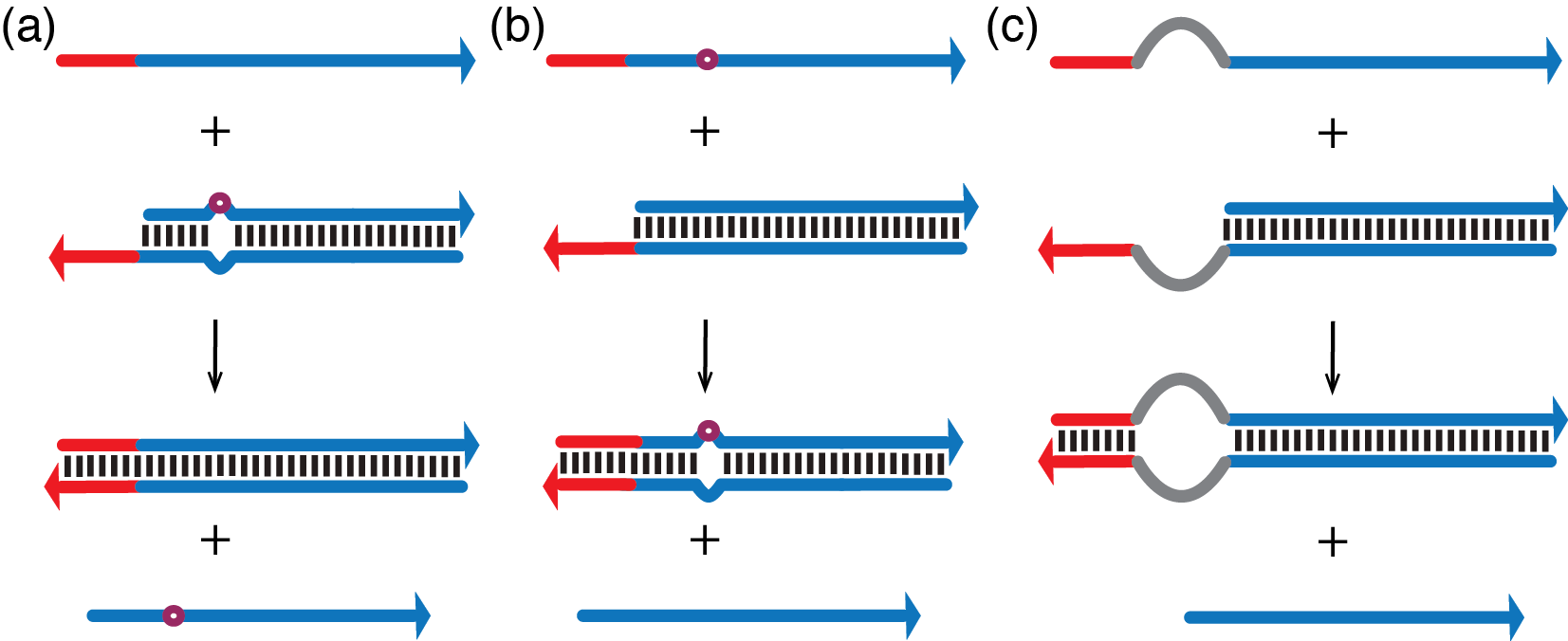}
  \caption{(a) Toehold-mediated strand displacement where the invading strand matches perfectly the substrate strand and the incumbent strand has a single mismatch with the substrate (mismatch repair). (b) Toehold-mediated strand displacement where the invader has a single mismatch with the substrate strand where the incumbent strand is complementary to the substrate (mismatch creation). (c) Remote toehold strand displacement with a non-complementary spacer region between the toehold and the displaced segment}
  \label{fig:mismatch}
\end{figure}

\subsection{Effects of mismatches}
For DNA, experimental studies have quantified the effects of mismatches in the branch migration region on the strand displacement kinetics.
In one recently studied case \cite{haley2018rational} (mismatch repair, Fig.~\ref{fig:mismatch}a), the invading strand corrects a mismatch that was present between the 
incumbent strand and the substrate. In a different study \cite{machinek2014programmable}, the other case was considered, in which the invader contains a nucleotide which is not 
complementary to the corresponding binding partner in the substrate strand (mismatch creation, 
Fig.~\ref{fig:mismatch}-b). In both cases, the change of rate was found to be highly dependent on the positions of the mismatch.

In the mismatch repair case, the displacement reaction rate has a non-monotonic dependence on the position of the mismatch between the substrate 
and the incumbent strand. Placing the mismatch three bases from the four-base toehold leads to approximately 75 times faster rate than for a system with no mismatches. The rate decreases if the mismatch is moved further away from the toehold. 
 This effect was rationalized through simulations, showing that the mismatch destabilizes the duplex, which increases probability that the invading strand will reach the position of the mismatch  and also increases the probability that the invader successfully completes the displacement. However, as the mismatch is moved further away from the the toehold, this effect becomes less pronounced, reaching a point where the rate is the same as when there is no mismatch present. 

Similarly, in the case where there is a mismatch between the invader and the substrate, the change of rate is highly dependent on the position of the mismatch. For toehold length 6, when the mismatch between the invader and the substrate is close to the toehold region, the rate of displacement is up to three orders of magnitude slower than if the mismatch is present at distant loci on substrate, away from the toehold. The effect is less pronounced for longer toeholds, which become stable enough that dissociation probability is low. The sensitivity to the position of the mismatch can be explained by the destabilization of the bound state in the case of the proximal mismatch, which can significantly decrease the probability of successful displacement. On the other hand, placing mismatch at the distal end of the of the displaced region has little effect on the probability of the displacement success. In the experiment 
\cite{machinek2014programmable}, the sequences were designed so that the final complex of the invader bound to the substrate have the same free energy for each placement of the mismatch, thus highlighting the observed changed of rate as a kinetic effect.

Finally, Genot et al.~\cite{genot2011remote} have also studied the case of a remote toehold (Fig.~\ref{fig:mismatch}c), where a spacer region of non-complementary bases was introduced between the invader toehold and the incumbent duplex. For toehold length that is well within the saturated regime (full binding of the invader to the toehold can be considered irreversible), it was shown that the rate of the displacement reaction is slowed down by increasing the spacer length, and was shown to slow down by two orders of magnitude for spacer length 17.

While no similar study of effects of mismatches has been done for RNA strand displacement with mismatches, we expect similar sensitivity of rate to position of mismatch position to apply to RNA as well (for toehold lengths shorter than the length at which the displacement saturates). 

\section{Molecular computing and engineered molecular machinery with strand displacement}
\label{sec:nanotech}
\begin{figure}
\centering
  \includegraphics[width=0.8\textwidth]{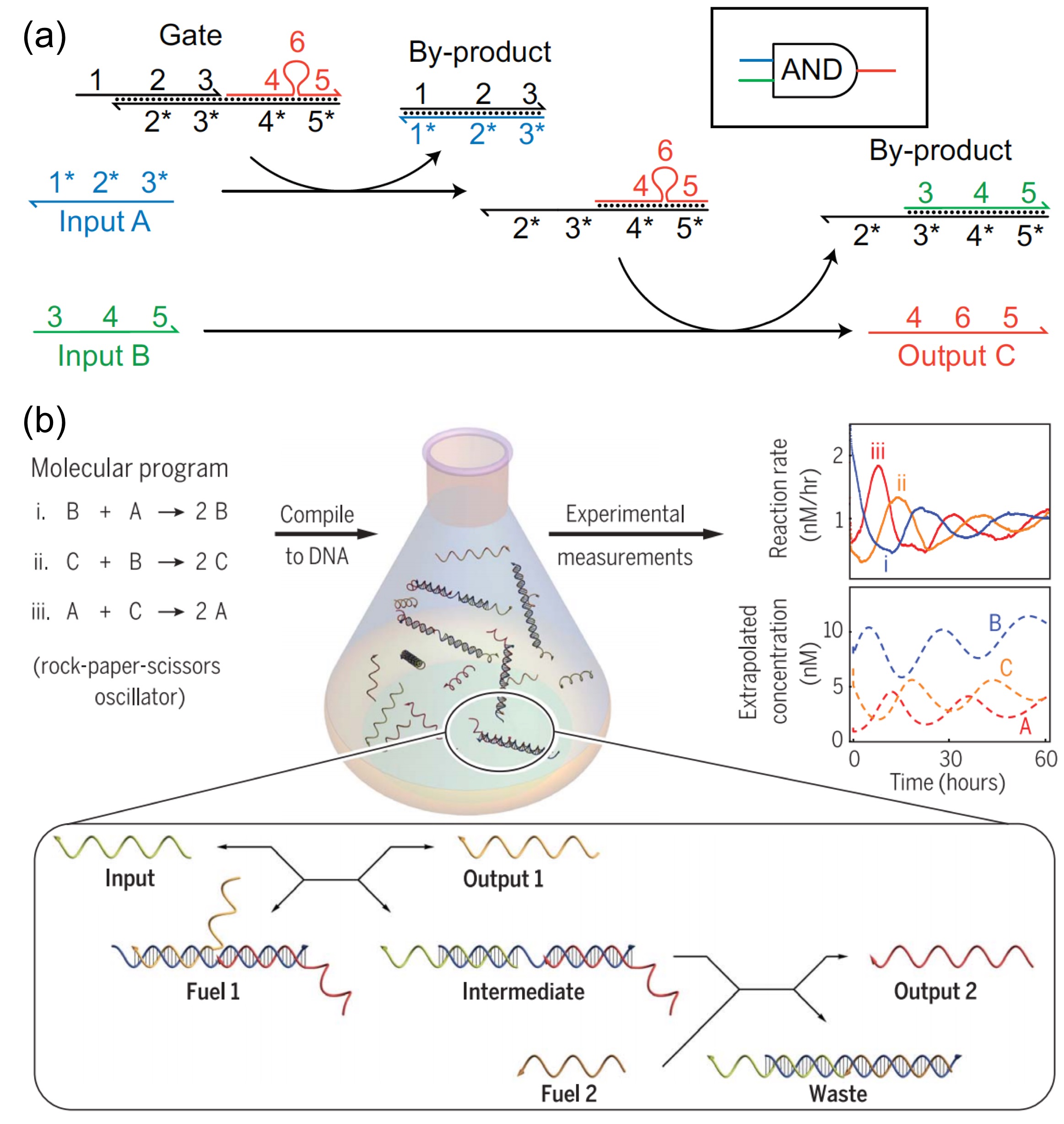}
  \caption{Implementation of DNA strand displacement for molecular computation (a) DNA strand displacement reaction based AND digital circuit. In the AND gate, if only one or none of the input A and B present, output C remains sequestered in the gate. Only if both A and B are present is C released through two sequential strand-displacement reactions. Input strand A will bind to the three-strand complex through toehold 1, and then B will displace the output red strand. (b) A molecular oscillator constructed from pure DNA strands. The DNA strand displacement is used to construct chemical reaction networks that realize "rock-paper-scissors" oscillation. The oscillations were observed in directly measured rates of the three implemented reaction modules and in the extrapolated concentrations of oligonucleotides representing formal species.}
  \label{fig:dnaexamples}
\end{figure}

Strand displacement reaction is a very powerful tool to control the molecular information flow and DNA/RNA's structure conformation, which can be used for molecular computing and function control of a designed machinery. 
The idea of molecular computing is to engineer a molecular system that can perform computational logic operations using molecules such as DNA or RNA in solution. These techniques are helpful in understanding basic cellular processes, constructing active nanotechnological devices, and providing new tools for medical applications such as diagnostics. Strand displacement reaction is a promising modular motif in this field because it has simple design rules and can be composed into a large scale computing framework. One displaced strand can trigger another set of strand displacement reactions, thus building up a network of interacting strands. In 2006, Takahashi et al.~\cite{takahashi2005chain} and Seelig et al.~\cite{seelig2006enzyme} constructed a set of DNA Boolean logic gates  via strand displacement reactions (Fig.~\ref{fig:dnaexamples}a). Seelig et al.~also developed a thresholding and an amplification process to solve the problems of leakage and signal normalization. The circuit is able to take miRNA as input, which shows promising applications in biotechnologies. Qian et al.~\cite{qian2011simple} developed a seesaw gate motif, which can represent digital OR and AND logic gates as strand displacement interactions between DNA strands. They successfully used a number of these strand-displacement-based gates to implement a complicated square rooter\cite{qian2011scaling} and an artificial neural network based on four fully connected neurons \cite{qian2011neural}. 
Soloveichik et al.~\cite{soloveichik2010dna} demonstrated that strand displacement reaction can be used to engineer arbitrary chemical reaction networks (Fig.~\ref{fig:dnaexamples}b). The idea has been experimentally demonstrated in DNA-based analog computation \cite{chen2013programmable} and pure DNA-based oscillator \cite{srinivas2017enzyme}. 

\begin{figure}
\centering
  \includegraphics[width=0.8\textwidth]{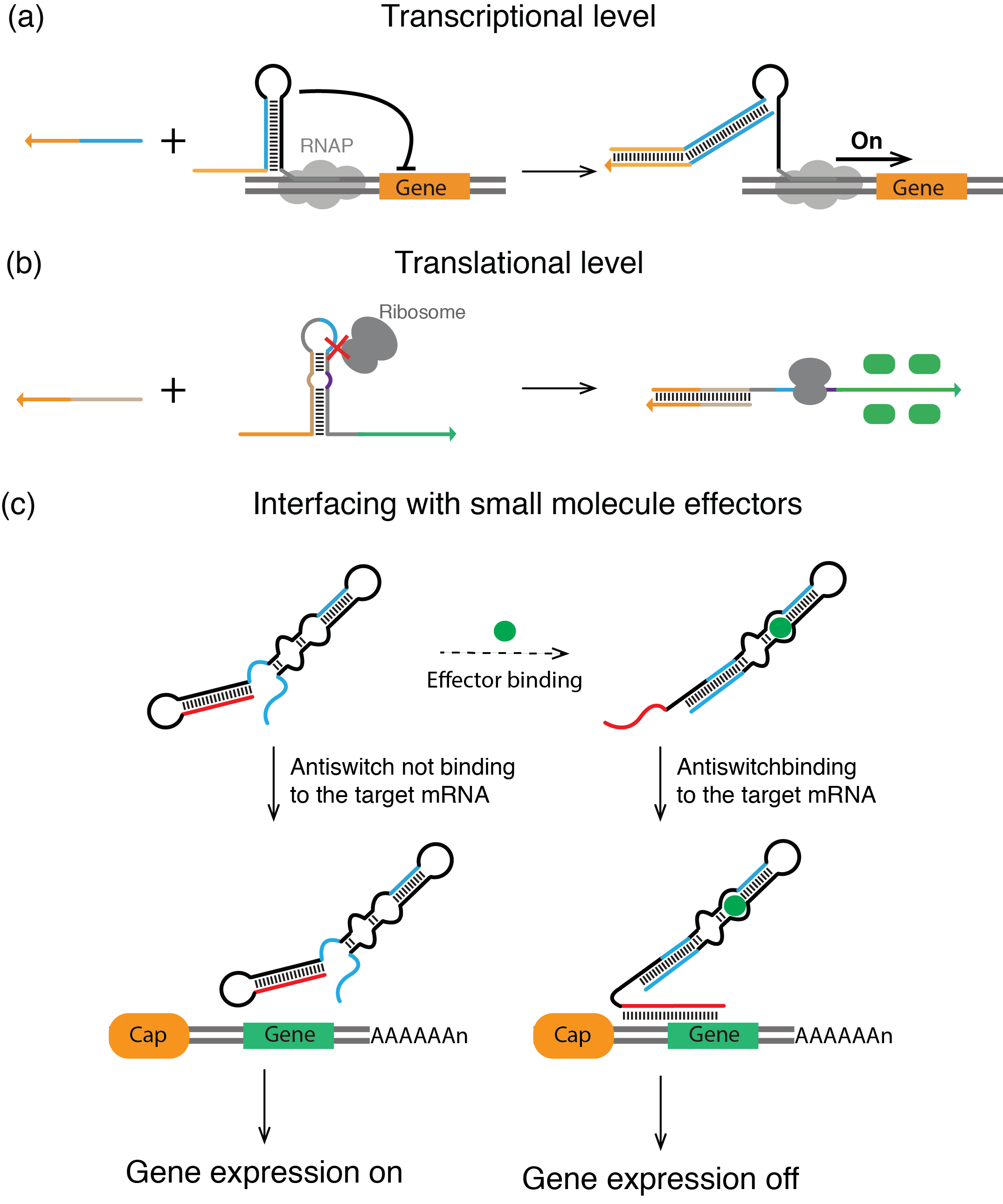}
  \caption{Implementation of RNA strand displacement for control of gene expression. (a) Transcription-level gene expression control based on the RNA strand displacement. The regulator small RNA binds to the terminator hairpin through the orange toehold region, opening the hairpin to allow access to the promoter region, activating transcription. (b) Translational level gene expression based on RNA strand displacement. Gene translation is inhibited because the RBS and start-codon are sequestered in a hairpin structure that is inaccessible by the ribosome. The regulatory RNA binds to the hairpin through the orange toehold to open the hairpin through strand displacement, exposing the RBS and start codon to initiate translation. (c) An engineered ligand-controlled riboregulator for eukaryotic gene expression based on the strand displacement. The antisense RNA is sequestered in a hairpin structure preventing interaction with the gene. After the binding of the effector, the toehold region is sufficiently strengthened to initiate strand displacement, opening the hairpin and releasing the antisense RNA, which binds to mRNA, inhibiting gene expression.}
  \label{fig:rnasynexamples}
\end{figure}

Other than molecular computing, the mechanism of strand displacement has also been applied to engineer a novel genetic motif to control gene expression. It's well known that RNA plays an important role in regulating cellular functions including splicing and editing of mRNA, modifying rRNA and regulation of gene expression as non-coding RNA.  The strand displacement can drive the conformation change of an RNA from one state to another, providing a feasible avenue to engineer molecular machines to control cellular functions.  For example, riboregulators are a type of small RNA that can both activate and repress bacterial gene expression by base paring with mRNAs and changing the secondary structure around translation initiation region, i.e. the ribosome biding site (RBS) and start codon. 
For example, Isaacs et al.~\cite{isaacs2004engineered} developed an artificial riboregulator which interacts with the YUNR (pYrimidine-Uracil-Nucleotide-puRine) sequence motif found in the 5' untranslated regions of mRNAs (Fig.~\ref{fig:rnasynexamples}a). The riboregulator uses the YUNR sequence as a toehold to displace a repressor sequence which blocks the RBS. 
 Mutalik et al.~\cite{mutalik2012rationally} also used YUNR interaction-driven strand displacement and successfully designed a tool for prediction and design of orthogonal RNA regulators. In 2014, by taking the full advantages of toehold mediated strand displacement, Green et al.~\cite{green2014toehold} designed a toehold switch that uses strand displacement mechanism to control the downstream gene expression (Fig.~\ref{fig:rnasynexamples}b ). In the toehold switch design, the RBS and starting codon were embedded into a hairpin structure to inhibit the ribosome's binding. The trigger RNA can interact with hairpin via the toehold region, then open the hairpin stepwise through strand displacement. After the strand displacement, the translation initiation region is exposed to the ribosome and the gene expression is subsequently activated. 
Chappell et al.~\cite{chappell2015creating} also used the strand displacement to engineer an RNA motif called STAR (small transcription activating RNA) that regulates gene expression at the transcription level (Fig.~\ref{fig:rnasynexamples}a). In the process of transcription in bacteria, the termination step requires the formation of a hairpin structure to push off the RNA polymerase and shut down the transcription. STAR binds to a transcribed RNA and opens the terminator hairpin to let the transcription continue, achieving up to a 9000-fold increase in gene expression. Other than the pure RNA-RNA interaction to control the cellular activity, RNA molecules can be engineered to interact with small molecules to activate the strand displacement reactions. For example, Bayer et al.~\cite{bayer2005programmable} designed a programmable ligand-controllable riboregulators which control the gene expression in eukaryotic cells. As shown in Fig.~\ref{fig:rnasynexamples}c, in the absence of the ligand, the anti-sense region will be sequestered in a stem region. However, in the presence of the small molecule effector, strand displacement will occur, exposing the anti-sense RNA, which will then interact with mRNA to inhibit its translation.

\section{Strand displacement in nature}

RNA structure's conformational change is thought to be fundamental to numerous biological process, including translational regulation \cite{putzer1992co,kumar1985plasmid}, protein synthesis \cite{wool1992ribotoxin}, and mRNA splicing \cite{lecuyer1994kinetics}. The structure change of RNA always involves two sets of interactions that are mutually exclusive, which indicates that the conformation change from one state to another one requires the breaking of one set and formation of the other one. The dissociation of a stable RNA helices happens with a large activation energy, and as a consequences the process becomes extremely slow at temperature below the melting temperature that disrupts the helices. Hence the spontaneous dissociation is not likely the driving force for the RNA structure switching because it will take significantly long time to complete. While helicases have been 
shown in many cases to provide an ATP-driven unwinding of DNA, RNA or RNA:DNA duplexes \cite{jankowsky2011rna}, it is also possible that in the cases where two RNA domains compete for binding to the same domain, the conformational change of the functional RNA undergoes non-ATP driven strand displacement that is faster and less temperature sensitive than the mechanism requiring full melted intermediate. 

In this section, we review processes that involve DNA or RNA interactions \textit{in vivo} that have been studied experimentally, and which have characteristics which suggest that a strand displacement reaction is involved. We include systems that involve exchange of base pairs between an invading single-stranded RNA domain and a substrate domain bound to an incumbent domain. The domains can be on different strand or can be located on the same strand. Such domains do not need to have perfect complementarity and can include mismatches. We consider only systems where the rearrangement or interactions can be explained without any active driving (such as helicase opening a stem). We note, however, that the explanations of the observed rearrangements of RNA/DNA strands as strand displacement systems is speculative in some cases, and it is possible it is actually driven by a different process.

\begin{figure}
\centering
  \includegraphics[width=0.8\textwidth]{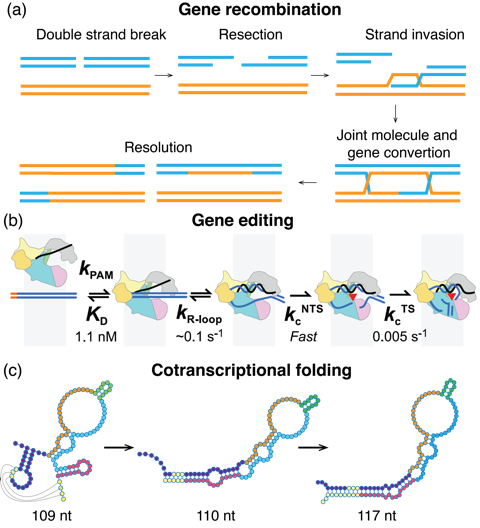}
  \caption{The mechanism of strand displacement in cellular functions. (a) The homologous gene recombination process. The homologous genetic recombination started when the DNA has a double stranded break, a set of enzymes, such as exonucleases, recombinases, polymerases and ligases, will drive the strand to invade a homologous DNA (orange) region to form a junction intermediate. After that, the Holliday junction can slide and/or extend along the joint molecule based on strand displacement reactions. Finally, alternative cutting and re-ligation of the double Holliday Junction intermediate can lead to a crossover\textemdash a swapping of DNA between the interacting chromatids. (b) The process of RNA guided genome cutting by CRISPR-Cas involves binding of the Cas protein, the R-loop formation and the DNA cleavage. The R-loop formation process contains the steps of RNA displace DNA duplex, which is the critical step allow specific genome targeting. (c) The cotranscriptional folding of {\it Escherichia coli} SRP RNA examined experimentally. In the RNA cotranscriptional folding process, newly transcribed fragments (yellow region) may act as toehold to initiate strand displacement, rearranging the conformation to the global minimal free-energy state.}
  
  \label{fig:rnaexamples}
\end{figure}

\textbf{The genetic recombination process} (Fig.~\ref{fig:rnaexamples}a) contains a branch migration step that allows the movement of exchange point (called the Holliday junction) between two homologous duplex DNAs \cite{panyutin1994kinetics,barzel2008finding}. If the branch point is flanked by DNA sequence homology, the Holliday junction can spontaneously migrate in either direction by the exchange of hydrogen bonds between the bases in homologous DNA strands. However, the dynamics of the process is different from the three-strand system described in Section \ref{sec:dispmechanism}, as it involves four strand exchange rather than three strand exchange, but it still involves base-pair exchange between different duplexes.

\textbf{The CRISPR-Cas} (Fig.~\ref{fig:rnaexamples}b) system has been recently discovered \cite{jinek2012programmable} to be a revolutionary tool for gene editing with wide applications \cite{hsu2014development}. The CRISPR-Cas system is composed of two parts, a nuclease that is used to cut the target DNA, and a programmable CRISPR RNA (crRNA) 
which provides the target sequence for nuclease activity. In the genome editing process, the guide RNA binds to the nuclease through a conserved hairpin structure to form an RNA-nuclease complex. The CRISPR-Cas complex then binds to a specific few nucleotide-long region (called PAM) on the DNA. Then, the spacer region of the guide RNA will lead the complex to the target genome region and disrupt the DNA:DNA duplex to form an RNA:DNA duplex, a process called 
R-loop propagation, leading to cleavage of both DNA strands. 
However, off-targeting is possible, as it has been shown that the CRISPR-Cas system can cut the dsDNA even if the crRNA is not fully complementary to the target DNA region
\cite{strohkendl2018kinetic,klein2018hybridization,boyle2017high}. Detailed understanding of the process of CRISPR-Cas binding is of great practical interest, especially with the goal of reducing/eliminating off-target effects. Experimental results on the CRISPR-Cas system point to the interpretation of the R-loop formation as a 
strand displacement process \cite{strohkendl2018kinetic,josephs2015structure}. Similar to the effects seen in \textit{in vitro} experiments \cite{machinek2014programmable}, proximal mismatches between the crRNA and the target DNA has a much stronger effect in the proximal region to the PAM binding site than at a distal position \cite{boyle2017high,szczelkun2014direct}.
As opposed to the DNA:DNA or RNA:RNA strand displacement, in this case, an RNA invader creates 
a DNA:RNA hybrid. Since DNA:RNA hybrid is more energetically stable than a DNA:DNA duplex of the same sequence \cite{sugimoto1995thermodynamic}, the corresponding free-energy landscape will not contain the plateau observed in Fig.~\ref{fig:dispkinetics}b and Fig.~\ref{fig:rnadisplacement}b, but instead will be decreasing as the invading RNA forms more bonds with the DNA strand. Instead of the toehold-binding, there will be a free-energy gain associated  
with binding of Cas protein to the PAM site. Furthermore, the free-energy landscape will be further affected by the presence of the Cas protein, which interacts with the DNA and RNA strand during the displacement process, likely affecting the height of the barrier as well as the slope of the branch-migration part of the free-energy profile.  

In fact, a recent work by Klein et al.~\cite{klein2018hybridization} has introduced a ``kinetic hybridization'' model of action of CRIPSR-Cas complexes to assess their off-targeting probability. Even though the authors do not make a connection to the existing strand-displacement literature and experiments from DNA and RNA nanotechnology, their proposed landscape for the R-loop formation share features with the free-energy landscapes from strand displacement simulations \cite{Displacement,machinek2014programmable}, including the penalty for mismatches between the invader and the substrate. The free-energy landscape used by Klein et al.~includes free-energy contribution of Cas binding to PAM region, as well as slope of the free-energy landscape of the free-energy landscape fitted to the experimental results for off targeting probability. Their model performs well in predicting cleavage probability when compared to the available experimental data for CRISPR-Cas9 and CRISPR-Cpf1 systems.



\textbf{Cotranscriptional folding of RNA} (Fig.~\ref{fig:rnaexamples}c): Another example of a natural system that likely involves strand displacement  is during the rearrangement of RNA during cotranscriptional folding. 
Nascent RNA folds as it is transcribed, with new base added approximately every $10-30\, {\rm ms}$ \cite{isambert2000modeling}. However, the initial folding of the RNA transcript does not necessarily correspond to the state with minimal free energy once the full sequence is produced. As the transcription continues, the newly transcribed fragments may interact with the older ones and rearrange the  conformation to change to its minimal state. Recent study by Yu et al.~\cite{angela2018computationally} suggests that strand displacement is involved in the structural arrangement during the contrascriptional folding. They studied folding of \textit{Escherichia coli} Signal Recognition Particle (SRP) RNA by using high-throughput RNA chemical probing that can capture the folding of RNA in a single nucleotide resolution \cite{watters2016cotranscriptional}. They found that the SRP RNA undergoes a dramatic conformation change from a multiple stem-loop structure into a long native helical structure as the length of the nascent transcript reaches around 110-111 nt. Based on their experimental data and molecular dynamic simulations,
they propose multiple possible pathways (one of them illustrated in Fig.~\ref{fig:rnaexamples}c) that involve strand-displacement mediated exchange between different complementary segments. In particular, the sensitivity of the rearrangement to the length of the nascent transcript is compatible with the toehold-mediated strand displacement kinetics, where the toehold length can change the kinetic rate by several orders of magnitude. Nascent transcript might for instance act as a toehold region for the displacement of the adjacent stem, as indicated in Fig.~\ref{fig:rnaexamples}. The strand displacement kinetics-mediated structural rearrangement is also compatible with the recent study of SRP folding \cite{fukuda2018alternative} by single-molecule force-pulling experiment. Yu et al.~note, however, that multiple rearrangement scenarios are possible, and further work is needed to understand the mechanism in more detail. In particular, different rearrangement pathways could be probed by introducing more mismatches in regions predicted to undergo strand displacement and compare the dynamics to predictions based on strand-displacement kinetics. 

We  expect that the strand displacement kinetics of rearrangement is applicable to other RNA folding systems, and in order to model accurately their folding dynamics, strand displacement cascades will need to be included in the modeling software as well \cite{sun2018predicting}. 

\textbf{Displacement of a competing binder}: Function of many ncRNAs in the cell involve binding to a complementary sequence. Examples include gene regulation by miRNA and siRNA \cite{he2004micrornas,filipowicz2005post}, where RNA binds to its target RNA and directs degradation. In the case of the target being blocked by a different partially complementary RNA through non-specific off-target binding, it is possible for the siRNA / miRNA to bind its target by displacing the previously bound RNA strand. Such a process is faster than spontaneous dissociation of the blocking strand. Similarly, it was previously discussed that displacement reactions occur {\it in vivo} in ribozyme-product complexes and it was shown {\it in vitro} that a single-stranded RNA can displace a previously cleaved strand which is bound to a ribozyme \cite{dissociation}. 

\textbf{Spliceosome cycle}. Large RNA-protein complexes, such as ribosome and spliceosome, require sequential binding of RNA and proteins. In the case of spliceosome, multiple of its components undergo rearrangement involving extensive base unwinding and change of binding domains over the course of splicing cycle \cite{wahl2009spliceosome}. While helicases are known to be involved in spliceosome remodeling during the cycle \cite{absmeier2016functions}, it has also been shown that designed oligonucleotides can unwind dsRNA in U4/U6 RNA complex by toehold-mediated strand displacement in the absence of a helicase \cite{rodgers2016multi}. It is thus possible that displacement might be involved in \textit{in vivo} remodeling of RNA-RNA interactions during the spliceosome cycle as well, and possibly in other intracellular protein-RNA complexes that involve rearrangements of the binding domains during assembly.

\textbf{Riboswitches}
Many bacterial mRNAs are regulated at the transcriptional and translational level by ligand-binding elements called riboswitch \cite{isaacs2006rna}. The function of some riboswitches require a dramatic conformation change upon binding to a ligand to be activated. For example, the natural adenine riboswitches for translational gene expression control has mutually exclusive pairing domains between the OFF and ON states. Lemay et al.~\cite{lemay2011comparative} found that the binding of adenine can push the structure transformation mRNA from OFF state to ON state. We suggest that due to the high activation energy for the hairpin to dissociate by itself, adenine may enhance the strength of a short toehold to initiate the strand displacement, switching the conformation. This mechanism is similar to the strategy Bayer et al.~\cite{bayer2005programmable} used to engineer ligand-responsive riboregulators to control gene expression in eukaryotic cells.

\section{Summary and outlook}
In this review, we introduced the basic principles of DNA and RNA strand displacement reaction, which has been used extensively in the fields of RNA and DNA nanotechnology, as well as synthetic biology, to construct dynamic systems. This reaction allows for an exchange of one single-stranded (incumbent) domain bound to a substrate by a competing (invading) domain. In particular, in the case of toehold-mediated strand displacement, the invading domain has extra base pairs (called the toehold) that it can form with the substrate. It has been shown experimentally for DNA that the rate of the strand displacement is highly dependent on the toehold region length, with saturation reached at 6-7 nucleotides.  In comparison, while the RNA strand displacement reactions have been successfully used in multiple designed systems \textit{in vivo} and \textit{in vitro}, there are currently no systematic measurements of the kinetic rates of the displacement as a function of varying toehold length and mismatches.

Strand displacement can explain multiple phenomena involving rearrangement and interactions of RNA in living systems. Speculations of rearrangement through branch migration can be found in the literature, and here we reviewed known experimental systems where RNA (DNA) strand displacement
is likely to play a role. In particular, we discussed DNA replication, CRISPR-Cas systems, contrascriptional folding of RNA, riboswitches, assembly of spliceosome, competing nonspecific binding of regulatory RNA, and binding of RNA to ribozyme. It is quite likely that there are other RNA-RNA interactions that involve strand displacement. In particular, strand displacement offers a method for rearrangement within a single RNA strand or between multiple RNA strands that does not require active unwinding involving helicases or other protein interactions. Recently developed techniques to study folding intermediates \cite{angela2018computationally,watters2016cotranscriptional} will soon enable us to study folding of many more types of RNA. Some of the already observed rearrangements likely does involve strand displacement, offering the possibility that it is a universal mechanism exploited by RNA molecules to reach their folded state via cotranscriptional folding pathway. 

More work remains to be done at the bioinformatic analysis, computational modeling, and experimental level. Efficient tools are needed to model cotranscriptional folding, and it is necessary for them to correctly incorporate the strand displacement kinetics. While strand displacement kinetics, including the effects of mismatches, has been studied for DNA, corresponding experiments for RNA still need to be carried out. If strand displacement is indeed a key process involved in folding of RNA, it should be possible to identify the conserved structural intermediates that are rearranged during strand displacement by studying covariation in homologous RNAs from different organisms. However, existing alignments are based on the final functional RNA structure, and it is possible that the signal of conserved intermediate structures is smeared out as these intermediates are not currently considered in alignments. Other bioinformatic methods, such as Transat \cite{wiebe2010transat}, have  been able to identify likely conserved helical segments in several RNAs, and will need to be combined with future high-throughput studies of RNA folding intermediates. 
More work is also needed on experimental characterization of RNA strand displacement kinetics in biological systems, as most studies have so far focused on DNA only.
Though simple conceptually, strand displacement may play an underappreciated role in biological systems, and understanding its mechanisms and impacts will broaden our understanding of cellular molecular biology, as well as lead to potential novel applications in diagnostics and therapeutics.

\section{Acknowledgments}
We thank Julius Lucks and Angela Yu for helpful discussions and comments and to Erik Poppleton for proofreading the manuscript. We are grateful to the participants of the Telluride Workshop on Challenges in RNA Structural Modeling and Design for inspiring discussions and feedback.
\bibliographystyle{unsrt}

\bibliography{ref}

\end{document}